# Close correlation between giant magnetostriction and the microstructure in Fe-Ga melt-spun ribbons

Likun Chen[1,2], Mitsutaka Sato[3], Jiahao Han[4], Rie Y. Umetsu[1,4]

[1] Initute for Material Research, Tohoku University, 2-1-1 Katahira, Aoba-ku, 980-8577 Sendai, Japan

[2] Mterials processing, Graduate School of Engineering, Tohoku University, 6-6 Aramaki-aza Aoba, Aoba-ku, 980-8579 Sendai, Japan

[3] Graduate School of Systems Science and Technology, Akita Prefectural University, 84-4 Tsuchiya Ebinokuchi, 015-0055 Yurihonjō, Japan

[4] Center for Science and Innovation in Spintronics, Tohoku University, 2-1-1 Katahira, Aoba-ku, 980-8577 Sendai, Japan

**Abstract**

Magnetoelastic anisotropy and ⟨100⟩ texture are crucial for promoting magnetostriction in Galfenol (Fe–Ga). Given that single-crystal Fe–Ga remains technically demanding and low magnetostriction of polycrystal, we explore melt-spun as an alternative, where rare-earth (RE) doping and cooling rates optimizing enable controllable <100> textured with required magnetoelastic anisotropy. $Fe_{81}Ga_{19}$ binary ribbons and RE-doped ribbons (0.2 at.% Pr, 1 at.% Pr and 1 at.% Ce) were fabricated at various cooling rates. Microstructural analyses reveal that RE elements preferentially dissolve into the matrix, while second-phase formation is suppressed at higher cooling rates. RE substitution increases the magnetostriction by enhancing magnetocrystalline distortion energy, while cooling rates act as an effective tuning knob to maximize the ⟨100⟩ texture. Notably, the decreased melting temperature associated with 1 at.% RE doping shifts the optimum texture to lower cooling rate compared with the binary alloy and 0.2 at.% RE doping sample. The magnetostriction as high as 688 ppm is achieved for the 1 at.% Ce doped ribbon fabricated at the speed of 1000 rotation per minute. These results demonstrate that RE-doped melt-spun ribbons are promising candidates for giant magnetostriction and establish a practical processing–texture–property guideline for designing highly magnetostrictive alloys.

## 1. Introduction

Magnetostrictive materials response as a length changing due to the external magnetic field, which leads to exchange between electrical and mechanical energies[1–10]. They have wide applications in sensors, actuators, biomedical materials and vibration power generation systems[3,11–18]. Terfenol ($TbFe_2$) attracted considerable attention due to its large magnetostriction, whereas the heavy involvement of rare-earth (RE) elements, as well as the insufficient ductility and toughness, makes it challenging to implement this material for many practical purposes[19–21]. In contrast, Galfenol (Fe-Ga) possesses not only reasonably large magnetostriction (400 ppm along the <100> direction in single crystals), but also strong ductility and low saturated magnetic field, thus being regarded as a more promising magnetostrictive materials[22–33]. To avoid high cost from synthesizing single crystals, polycrystalline Galfenol is favored in large-scale productions, but there arises a critical problem that the magnetostrictrion is weakened due to the diminishing magnetoelastic anisotropy of the less crystallized structure[25,32,34–38].

Substitutional alloying has been widely explored to overcome the issue of limited magnetostriction in polycrystalline Fe–Ga. For example, partial replacement of Ga with Ge or Al can enhance the magnetostriction; however, the resulting values are still lower than those of binary Fe–Ga single crystals[39–41]. Additions such as B element have also been proposed because of its ability to occupy interstitial sites in the lattice and Sn can induce the $D0_3$ phase stability, although all these approaches do not effectively control crystallographic texture and therefore cannot substantially increase the magnetoelastic anisotropy[41–43]. Similarly, annealing for removing stress can improve magnetostriction to some extent, yet the achievable magnetostriction is still lower than the target level required for high-performance applications[44,45].

More recently, it has been reported that dilute RE doping (typically less than 1 at.%) can further enhance magnetostriction, which has been attributed to the large magnetocrystalline distortion energy of RE elements and their tendency to substitute specific Fe sites that preserve the underlying chemical ordering[33,36,38,46–52]. Nevertheless, because of the substantial atomic-size mismatch between RE and the Fe or Ga atoms, the solubility of RE is severely limited, usually below ~0.4 at.% for RE elements and even below ~0.1 at.% for La[36,53–57]. The existence of secondary phases containing the excess RE elements intrinsically restricts the maximum attainable enhancement and also suppresses the mechanical properties[33]. Moreover, higher RE additions readily lead to the formation of RE-rich secondary phases; when these phases segregate to grain boundaries, they hinder magnetostrictive elongation and consequently reduce the overall magnetostriction[33,48–52,56].

To address the above problem, melt-spun, a method that controls the crystalline texture via tunable cooling rates (i. e., rotating speeds), has been applied to lift the magnetostriction of Fe-Ga[36,43,53,54,58–72]. This method also increases the RE solubility in Fe-Ga than before, so that the large negative quadrupole moments of RE would be more useful to promote the tetragonal deformation of Fe-Ga[36,47,54,68]. As a result, the magnetostriction of polycrystalline Fe-Ga has been increased from the original value of 40 ppm to 500~1000 ppm by melt-spun[36,53,54,58,69–72]. Nevertheless, the large discrepancy in the reported magnetostriction values and unclear correlation between fabrication process, crystal texture and magnetostriction poses fundamental obstacles in elucidating the enhancing mechanism.

In this work, we find that tailoring the cooling rate (i.e. rotating speed) during melt spinning enables undercooling-engineered ⟨100⟩ texture and giant magnetostriction (up to 688 ppm) in RE-doped $Fe_{81}Ga_{19}$ ribbons[73–76]. Specifically, the cooling rate was optimized for each composition to maximize ⟨100⟩ texture development. With increasing cooling rate, RE-rich secondary phases are progressively suppressed, and RE atoms preferentially dissolve into the matrix[47–53,55,67–71]. Texture was evaluated by electron backscatter diffraction (EBSD) on the ribbon both free face and cross section, which reliably represents the overall grain alignment to determine the crystal orientation arrangement of the whole samples as the key accomplishment[23–25]. For each composition, the sample exhibiting the highly preferred ⟨100⟩ texture also shows the highest saturation magnetostriction[33,47,60,62]. Overall, this work establishes a consistent processing–texture–property relationship, explains the large performance scatter often observed among samples. These findings identify RE-doped melt-spun Fe–Ga ribbons are promising candidates for giant magnetostriction and provide practical guidelines for synthesizing highly magnetostrictive Fe–Ga-based materials.

## 2. Results

### 2.1. Microstructure and crystal characterization

Figures 1 (a)~(c) and (d)~(f) show images of microstructure observation by scanning electron microscope (SEM) of melt-spun ribbons fabricated in different rotating speeds at the free face for $Fe_{81}Ga_{19}$ and the 1 at.% Ce doped samples (denoted as + Ce), respectively. In sample fabrication, there are two kinds of surfaces in contact with the high-speed roller (roll face) and with the atmosphere (free face). The microstructures of $Fe_{81}Ga_{19}$ observed at roll face and free face in three different rotating speeds in same enlargement are shown in figures S1 (a)~(c). Grains on the roll face are smaller than those on the free face because of the higher cooling rate at the roller interface. In both the roll face and free face of the $Fe_{81}Ga_{19}$ binary alloy, the grain size decreases with increasing wheel speed, reflecting the shorter growth time caused by higher cooling rates. In RE elements doping samples such as + Ce, white-ball-like secondary phases exist in the free face, the number of them decreases with rotating speed increasing as observed at figures 1 (d)~(f). In addition, a similar tendency is observed for the other RE-doped alloys see figures S1 (d)~(f) of 1 at.% Pr doped samples (denoted as + Pr). Figures 1 (g)~(j) indicate the areas evaluating the compositions by energy dispersive X-ray spectroscopy (EDX) at the free face for $Fe_{81}Ga_{19}$, 0.2 at.% Pr doped samples (denoted as + $Pr_{0.2}$), + Pr, respectively, and at the role face for + Pr, here, all of the samples were fabricated by the same rotating speed of 1 krpm. The EDX analyses were performed at full map and each area and point, and the numerical values are listed in Table 1. From the results, the binary system has homogeneous composition of $Fe_{81}Ga_{19}$ and RE doped systems exhibit secondary phases containing high RE composition caused by the low solubility limitation. The RE rich phases even exist in the + $Pr_{0.2}$ sample as shown in figures 1 (h)~(j) and Table 1.

To clarify the distribution of RE elements at different cooling rates of 1, 2 and 3 krpm, X-ray diffraction (XRD) of each sample was measured and the patterns are plotted in figures 2 (a) ~ (d).

Here, those bulk alloys with the same composition are also indicated in each figure. In the $Fe_{81}Ga_{19}$ basic system, four characteristic peaks are assigned to the disordered A2 phase (figure 2 (a))[38,77]. The XRD patterns of RE doped samples possess the same four main peaks, together with additional reflections for the +Pr samples fabricated at 1 and 2 krpm and for all +Ce samples (figures 2 (c), (d), figures S2 (a)~(c)). No extra peaks are detected for $+Pr_{0.2}$ although microscale RE-rich precipitates are observed (figure 2 (b)). When the Pr content is increased to 1 at.% (+Pr), a larger amount of secondary phases is detected in XRD patterns, as evidenced by additional peaks. However, the intensities of these peaks decrease and even vanish at higher cooling rates (figure 2 (c)). A fraction of the RE-rich phases may also be confined to grain boundaries, which would diminish or eliminate their diffraction peaks in the patterns of the bulk. Furthermore, because Ce has a slightly larger atomic radius (consistent with the lanthanide-contraction trend), less Ce dissolves into the matrix, resulting in more numerous and more intense extra peaks (figure 2 (d)).

By combining phase diagram[77], EDX compositions, and XRD pattern, we attribute the additional diffraction peaks mainly to $RGa_2$ and $R_2Fe_{17}$ (R = RE). With increasing cooling rate, the peaks corresponding to $RGa_2$ are more significantly suppressed or even disappear, whereas the $R_2Fe_{17}$ peaks remain, suggesting that RE atoms in $RGa_2$ are more likely to dissolve into the matrix than those in $R_2Fe_{17}$.

By overlapping and normalizing the intensity of 200 peak, we can see that the position of each peak is shifted slightly depending on the different rotating speeds, indicating the variation of the lattice parameters in figure 2 (e). The corresponding lattice parameters were refined and plotted in figure 2 (f) [78]. The lattice parameter of the $Fe_{81}Ga_{19}$ bulk sample is close to the dashed line representing the reported value[79]. RE doping leads to an increasing the lattice parameter owing to the larger atomic radius of the RE elements. On the other hand, all ribbon samples exhibit smaller lattice parameters than that of each bulk sample, consistent with residual stresses introduced during rapid solidification. At higher cooling rates, larger residual stresses would be introduced, which would drive a monotonic decrease in the lattice parameter from 1 to 3 krpm of the rotation speeds for both $Fe_{81}Ga_{19}$ and $+Pr_{0.2}$.

Notably, when RE-rich second phases segregate along grain boundaries[33,38,48–53,67–71], they can exert a compressive constraint on the matrix, resulting in a reduced apparent lattice parameter, which explains the lower lattice parameters of the +Ce and +Pr samples compared with $Fe_{81}Ga_{19}$ in 1 and 2 krpm. In contrast, as the cooling rate increases, more large RE atoms enter the matrix under non-equilibrium solidification, leading to an expansion of the crystal lattice and, consequently, an increase in the lattice parameter for the +Pr and +Ce ribbons. In summary, RE atoms preferentially form grain-boundary secondary phases at low cooling rates, whereas at higher cooling rates they are more likely to be incorporated into the matrix, which is reflected in the systematic evolution of the lattice parameter.

## 2.2 Magnetization and Curie temperature

The magnetization curve (*M-H*) of the $Fe_{81}Ga_{19}$ 1krpm ribbon sample was measured at 5 K in magnetic fields ranging from −5 to 5 T and shown in figure 3 (a)[38]. The hysteresis loop exhibits a very low coercive field and a high saturation magnetization of 178.3 $A \cdot m^2 \cdot kg^{-1}$. Both the bulk and ribbon samples exhibit high saturation magnetization, $M_s$. However, $M_s$ decreases upon doping with non-ferromagnetic RE elements (figure 3 (b)). The bulk sample shows a higher $M_s$ than the ribbons, which would be attributed to its smaller oxidized surface fraction[38]. The fabrication conditions also affect $M_s$ in the $Fe_{81}Ga_{19}$ binary alloy, $M_s$ increases with increasing the cooling rate, which can be ascribed to the suppression of Ga-rich short-range order and the correspondingly higher $M_s$ of the disordered A2 phase[62,77]. In contrast, for the RE-doped samples, $M_s$ decreases with increasing the cooling rate. As discussed in Section 2.1 and in previous reports[47], higher cooling rates promote the incorporation of RE atoms into the matrix, as confirmed by the result of SEM–EDX and XRD, and this increased amount of non-ferromagnetic RE in the matrix leads to the reduction of $M_s$.

Figure 3 (c) indicates differential scanning calorimetry (DSC) heating curve and temperature dependence of magnetization (*M-T*) of $Fe_{81}Ga_{19}$. The endothermic peak observed around 970 K in the DSC curve corresponds to the ferromagnetic–paramagnetic transition and is identified as the Curie temperature, $T_C$, in agreement with the *M–T* measurements[38]. The DSC heating curves for all samples are presented in figures S4 (a)~(d). Similar to the behavior of $M_s$, $T_C$ decreases when non-ferromagnetic RE elements are doped into the matrix (figure 3 (d)). In addition, the specimens with larger $M_s$ tend to indicate higher $T_C$. However, the trend does not follow the relation as in the change of the rotation speeds. That is, $M_s$ decreases with increasing the rotation speeds, whereas $T_C$ tends to increase. Various factors may affect $T_C$ at elevating temperatures but the absolute differences among processing conditions are relatively small.

## 2.3 Correlation of texture and magnetostriction

For better understanding the correlation between texture and the magnetostriction, using EBSD to observe the crystal orientation of free face was reported by the group of C. Saito[60], but only free face observing is hard to represent the whole crystal arrangement of ribbon samples. Without only observing the microstructure in free face, the cross section of the ribbons is also observed in the present investigations. EBSD maps, polar figures, and inverse polar figures observed in both free face and cross section in A1 direction are indicated in figure 4. Here, the A1 direction is parallel to the direction used for the parallel magnetostriction measurements. Figures 4 (a)~(c) are those for $Fe_{81}Ga_{19}$ ribbons fabricated by 1, 2, and 3 krpm rotating speeds, respectively, and figures 4 (d)~(f) are for +$Pr_{0.2}$-2 krpm, +Pr-1 krpm and +Ce-1 krpm, respectively, in where the most preferable texture is obtained in each system. With increasing cooling rate, both the ribbon thickness and grain size decrease, and the texture is markedly modified. From comparison of the EBSD map between the free face and the cross section, it is found that the discussion of the grain size may be reasonable, but the texture of both is completely different. It should be noted that the texture of the surface does not always correspond to that of the cross section which represents mostly the whole ribbon.

In the $Fe_{81}Ga_{19}$ binary alloy, the 1 krpm sample exhibits coarse equiaxed grains with an average size of ~50 μm, and most grains are oriented randomly (figure 4 (a)). As the rotating speed increases, the size of crystal in free face to around 10 μm and the 2 krpm sample developed a columnar structure with grains in the cross section. The crystal arrangement aligns near the ⟨100⟩ direction in both faces of the 2 krpm, whereas the columnar grains in the 3 krpm sample become less well aligned though the free face of them are similar. This phenomenon determines observing the free face of ribbon is not enough whereas the cross section is the most representative face to analyzing the alignment of the whole ribbon samples.

Similar behavior is observed in the RE-doped specimens: each composition shows an optimal ⟨100⟩ alignment, which occurs at 2 krpm for $+Pr_{0.2}$ and at 1 krpm for +Pr and +Ce as shown in figures 4 (d)~(f), figures S5 (a)~(c). The degree of alignment was quantified by converting the deviation of each grain from the ⟨100⟩, ⟨101⟩, and ⟨111⟩ directions, where 0°, 45°, and 54.73° correspond to exact ⟨100⟩, ⟨101⟩, and ⟨111⟩ directions, respectively. In the optimally textured $Fe_{81}Ga_{19}$–2 krpm sample, many grains lie close to the ⟨100⟩ direction, giving the most preferable ⟨100⟩ texture (right figure in 4(b)). Samples with the lowest <100> deviation angle exhibit the highest fraction of ⟨100⟩-oriented grains, whereas the other samples show a more random texture superimposed on a weaker ⟨100⟩ component (Fig. 4(b)).

At $Fe_{81}Ga_{19}$–3 krpm, however, all compositions exhibit a much more random texture, demonstrating the dominant influence of nucleation at very high cooling rates. In $Fe_{81}Ga_{19}$–2 krpm some columnar grains extend across the entire ribbon thickness, whereas such through-thickness columnar grains are absent at 3 krpm. The same tendency is also observed in the RE-doped samples that the columnar crystal cannot grow across the whole cross section in higher cooling rates. This can be speculated that the nucleation not only happened in the interface between molten metal and rolling wheel but also happened inside of it.

As the group of C. Jiang previously reported in how cooling rates affecting the crystal orientation and magnetostriction, the magnetostriction is evaluated and greatly correlated with texture which was evaluated in parallel directions to the magnetic field by strain gauge[25,80–83]. The magnetostriction curves, $\lambda$–$H$, for all samples are shown in the Supporting Information (Fig. S6). All ribbon samples exhibit larger saturation magnetostriction than that of the bulk sample, and all RE-doped ribbons show enhanced magnetostriction compared with the pristine $Fe_{81}Ga_{19}$ ribbon, except for $+Pr_{0.2}$–1 krpm[38]. This overall trend demonstrates the strong enhancement arising from both RE doping and melt spinning. In combination with the EBSD results, the anomalously low value of $+Pr_{0.2}$–1 krpm is attributed to the fact that most grains are oriented far away from the ⟨100⟩ direction.

The value of the magnetostriction is closely correlated with the deviation angle to the <100> crystal orientation. Among all compositions and processing conditions, the +Ce–1 krpm ribbon exhibits the highest magnetostriction of 688 ppm, followed by $+Pr_{0.2}$–2 krpm with 541 ppm and +Pr–1 krpm with 402 ppm. In the $Fe_{81}Ga_{19}$ binary system, the 2 krpm ribbon shows the largest magnetostriction of 173 ppm among all $Fe_{81}Ga_{19}$ samples as shown in figure 6 (d).

Also, the magnetostrictive value of each composition is also different. In the Pr-doped system, 0.2 at.% Pr was more favorable than 1 at.% Pr because excessive Pr promoted second-phase formation and suppressed magnetostriction, even though the 1 at.% Pr ribbon processed at 1 krpm showed the highest ⟨100⟩ texture within that composition series[56,57]. In the Ce-doped system, although the bulk compositional trend suggested an optimum around 0.8 at.% Ce, the highest magnetostriction in melt-spun ribbons was achieved for 1 at.% Ce processed at 1 krpm, indicating that the apparent optimum composition can shift under specific cooling-rate conditions through the combined effects of texture and solubility[50,52]. Also, +Ce-1 krpm shows lower average angle than +$Pr_{0.2}$-2 krpm. Under the optimized condition and closer average crystal orientation to <100>, a maximum saturation magnetostriction of 688 ppm was obtained.

Taken together, for all ribbon samples, higher saturation magnetostriction corresponds to a smaller <100> deviation angle, confirming that a preferable ⟨100⟩ columnar texture is the primary factor governing the magnetostrictive response, while the solubility of the RE element exerts a secondary influence. Thus, by optimizing the cooling rate during melt-spun, <100>-textured ribbons with giant magnetostriction can be reproducibly obtained through the robust coupling between crystallographic texture and magnetostriction.

**2.4 Model of texture arrangement**

One question still remains why the most preferable ⟨100⟩ alignment is obtained at 2 krpm for $Fe_{81}Ga_{19}$ and +$Pr_{0.2}$ but at 1 krpm for +Ce and +Pr. The melting temperatures of the alloys were evaluated by DSC and indicated in figure 5 (c) and figure S5 (d). Considering the presence of RE-rich secondary phases revealed by SEM–EDX and XRD, the firstly appeared DSC peak in heating process is attributed to the melting temperature of the RE-rich phase, whereas the second peak corresponds to that of the Fe–Ga-based matrix. $Fe_{81}Ga_{19}$ and +$Pr_{0.2}$ exhibit nearly identical melting temperatures of around 1670 K, while the melting temperatures of the +Pr and +Ce alloys with 1 at.% RE are reduced to around 1560 K as shown in figure S5 (d). The correlation between undercooling ($\triangle T = T_{\mathrm{m}} - T$) and the velocity of the moving boundary to the undercooling is expressed as follows.

$$V_{\mathrm{hkl}} = \mu \cdot \triangle T \tag{1}$$

where, $T_m$ is the melting temperature and $\mu$ is the kinetic constant of the crystal-melt interface that is different from each crystal orientation <hkl>. D.Y. Sun *et al.* [84] reported that $\mu_{<100>}$ has the highest value which supposed <100> has the highest growing speed, so it has the most preference in fast cooling rates than other crystal orientations. With spite of the deviations from the hard-sphere system for specific systems like Fe, the $\mu$ is expressed by M. Amini *et al.* [76]as

$$\mu = C_{\mathrm{hkl}}\sqrt{R/MT_{\mathrm{m}}} \tag{2}$$

here, $R$ is the gas constant, $M$ is the molar mass and $C_{\mathrm{hkl}}$ is a constant that depends upon the crystal orientation <hkl> and $C_{100}$ has the highest value. Considering higher cooling rates, <100> crystal will occupy larger area whereas the nucleation starts inside the molten metal and prohibits the growth, the nucleation rates can be described by next relation [85,86]

$$I = (N_{\mathrm{s}} - N)K_1 \exp\left(-\frac{\Delta G_{\mathrm{cr}}}{k_{\mathrm{B}}T}\right) \tag{3}$$

where $I$ is the nucleation rate, $N_{\mathrm{s}}$ is the number of heterogeneous substrates originally available per unit volume, $N$ is the number of particles that have already nucleated. $K_1$ is a constant, $k_B$ is the Boltzmann constant, $T$ is the present temperature and $\Delta G_{\mathrm{cr}}$ is excess free energy of the critical nucleus. The competition happened in the beginning of the nucleation so it should be $N_s >> N$, so the equation (3) can be simplified as[85,86]

$$I = N_s K_1 \exp\left(-\frac{\Delta G_{\mathrm{cr}}}{k_{\mathrm{B}} T}\right) \tag{4}$$

The excess free energy of the critical nucleus ($\Delta G_{cr}$), which the crystallization must overcome, described as[85,86]

$$\Delta G_{\mathrm{cr}} = \frac{16\pi\gamma^3 T_{\mathrm{m}}^2}{3\Delta H_{\mathrm{m}}^2 \Delta T^2} \tag{5}$$

where $\gamma$ is liquid-solid interface energy and $\Delta H_m^2$ is enthalpy of melting. Both parameters are assumed to remain nearly constant within the present composition range, whereas the undercooling $\Delta T$ is strongly affected by the cooling rate during melt-spun. $\Delta G_{cr}$ decreases at higher cooling rates because $\Delta T$ becomes to be larger. This suggests that an ideal degree of undercooling exists that is suitable for crystal growth while suppressing nucleation in the molten metal except at the interface. Accordingly, the cooling rate (i.e., the rotating speed in our experiment) was varied among 1, 2 and 3 krpm in order to identify the optimal condition for fabricating textured samples, as schematically sketched in figure 4 and S5.

Under this approximation, a decrease in $T_m$ with RE elements doping leads to a decrease of $\Delta G_{cr}$ and rising on the nucleation rates in molten metal. Thus, to maintain a high growth rate along the ⟨100⟩ direction and suppress the nucleation inside of the molten metal, $\Delta T$ was increased with increasing rotating speed, which support the preferable <100> crystal orientation in $Fe_{81}Ga_{19}$ and $+Pr_{0.2}$ system in 2 krpm but in +Ce and +Pr in 1 krpm.

According to the equations (4) and (5), an increase in $\Delta T$ reduces the nucleation barrier, thereby increasing the nucleation rate per unit volume and competing with, or even suppressing the directional growth of ⟨100⟩-oriented grains[85,86].

## 3. Conclusion

In summary, this work demonstrates that melt spinning is an effective route for enhancing the magnetostrictive performance of $Fe_{81}Ga_{19}$, $(Fe_{81}Ga_{19})_{99.8}Pr_{0.2}$, $(Fe_{81}Ga_{19})_{99}Pr_1$ and $(Fe_{81}Ga_{19})_{99}Ce_1$ ribbons through cooling-rate-controlled microstructure engineering. By systematically varying the melt-spinning condition and correlating SEM–EDX, XRD, EBSD, and magnetic/magnetostrictive measurements, a clear processing–microstructure–property relationship was established. The results show that under nonequilibrium solidification, cooling rate strongly influences both rare-earth solubility and crystallographic texture, thereby governing the magnetostrictive property of the ribbons.

Regarding the magnetostrictive property of RE-doped Fe-Ga melt-spun ribbon is primarily controlled by the formation of a pronounced ⟨100⟩ columnar texture on the ribbon cross section, with RE solubility acting as a secondary but coupled factor through the suppression or formation of RE-rich secondary phases. For Pr doping, 0.2 at.% Pr provided a more favorable balance than 1 at.% Pr, since excessive Pr promoted secondary phases and degraded magnetostriction, even when strong ⟨100⟩ texture was developed. For Ce doping, the average crystal orientation of 1 at.% Ce doping sample is closer to <100> and approximate to the solubility limitation. Consequently, a maximum saturation magnetostriction of 688 ppm was achieved for the 1 at.% Ce ribbon processed at 1 krpm.

The cooling-rate dependence of the preferred ⟨100⟩ texture can be rationalized by a crystal growth–nucleation competition system. Increased undercooling enhances the growth advantage of the ⟨100⟩ orientation, but simultaneously promotes nucleation in the melt, limiting the development of a highly aligned columnar structure. This balance explains why the highest ⟨100⟩ texture emerges only within a specific cooling-rate window and why the optimum processing condition varies with composition.

These findings identify cooling-rate engineering during melt spinning as a powerful strategy for tailoring texture–solubility coupling in Fe–Ga–RE polycrystals and provide a practical design framework for scalable high-performance magnetostrictive materials and devices.

## 4. Experimental method

Precursor ingots $Fe_{81}Ga_{19}$, $(Fe_{81}Ga_{19})_{99.8}Pr_{0.2}$, $(Fe_{81}Ga_{19})_{99}Pr_1$ and $(Fe_{81}Ga_{19})_{99}Ce_1$ (denoted as +0.2Pr, +Pr and +Ce, in whole of the present report, respectively) were prepared from high purity 99.99 % Fe and Ga, and 99.9% Ce and Pr by high frequency furnace. One part of them were applied into heat treatment in 1373 K for 3 days. Other parts were prepared for ribbon samples by melt-spun. For better understanding the influence of cooling rates on the microstructure, three different rotating speeds of 1, 2 and 3 krpm (10.5, 21.0 and 31.5 m/s, respectively) have been applied in all these four systems. The image of melt-spun method and the efficiency of cooling rates are illustrated in figure 6 (a). Width and thickness of ribbons fabricated in the speed of 1 krpm are about 5 mm and 100 μm, and they are decreased by rotating speed increasing.

XRD was carried out with Co-$K_\alpha$ radiation source. PowderCell 2.4 software was used to refine the lattice parameters. Microstructure and composition were evaluated with SEM-EDX. The crystal orientations of free face and cross section were evaluated by EBSD. Polar figure and inverse polar figure were further analyzed. In order to get clear microstructure data of the cross section, a holder would hold the ribbon sample and compressed into the resin. After grinding up to 2400 grit Sandpaper, ion milling was taken to polish the surface.

Magnetization (*M-H*) curve was measured by SQUID magnetometer in applying magnetic field up to 5 T. Full loop of *M-H* curve was only applied in $Fe_{81}Ga_{19}$ sample to evaluate the coercivity. The Curie temperature was investigated mainly by DSC experiments because the maximum elevating temperature is higher than that for the limit of *M-T* measurements by VSM option in the physical property measurement system (PPMS, Quantum Design Ltd.). Melting temperature of the mother alloys was also evaluated by DSC in the temperature range up to 1773 K. Strain gauge was stuck onto the roll face of the ribbon specimens in order to investigate the magnetostriction. The magnetic field from -1.5 T to 1.5 T was applied by electromagnet and the magnetostriction in parallel and perpendicular directions to the magnetic fields was evaluated as shown in figure 6 (b).

**Acknowledgement**

The authors express sincere thanks to Assistant Professor H. Yim, and Emeritus Professors Y. Kawazoe and S. Suzuki in Tohoku University for their fruitful discussions. They also thanks cooperation to Ms. Qian Chen for her graphical drawing works. We acknowledge the support from the Cooperative Research and Development Center for Advanced Materials of IMR, Tohoku University. This work is supported by Grant-in-Aid for Challenging Research (Exploratory), MEXT 22K18879, Japan.

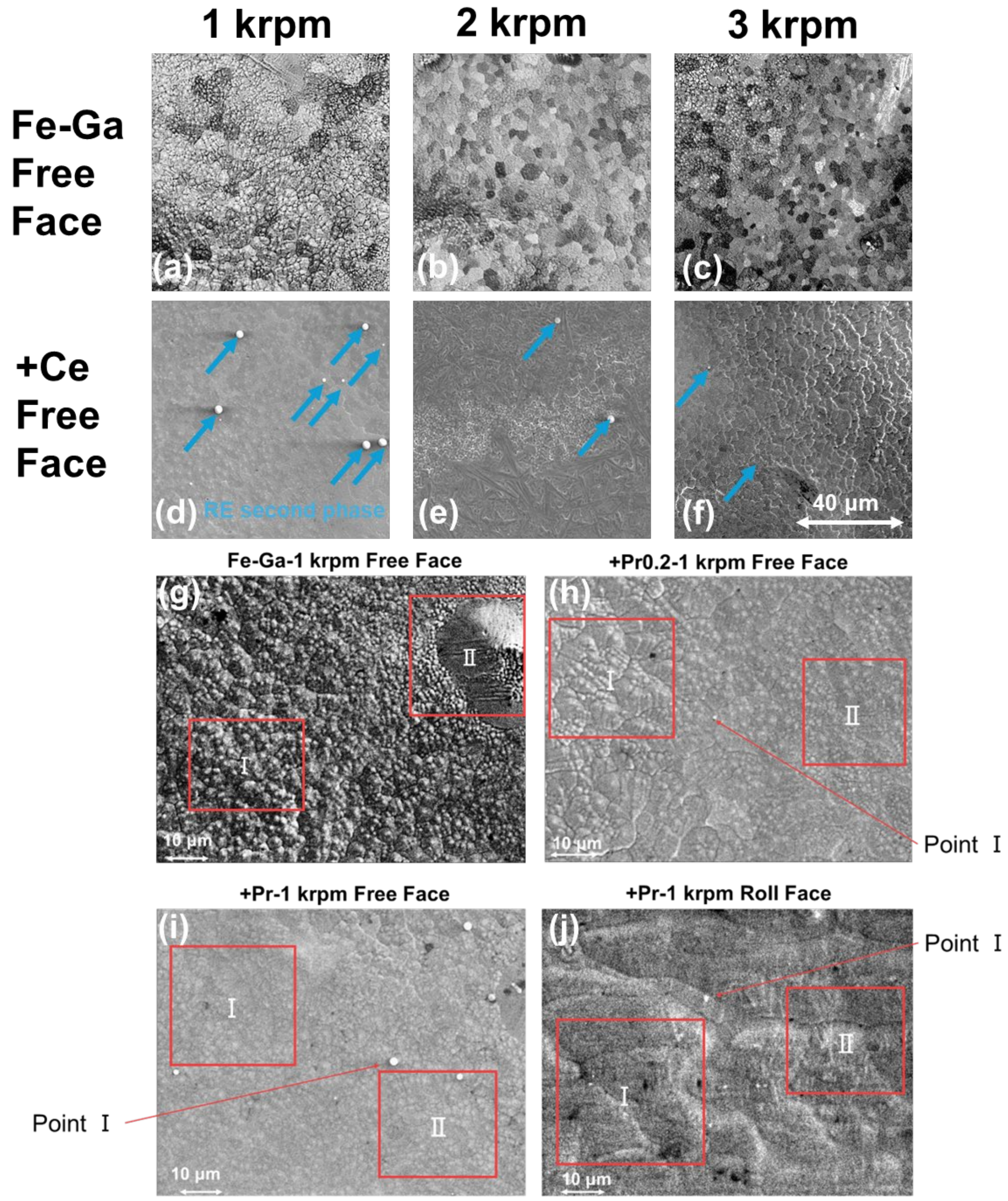

Figure 1 SEM images observed at the free face of $Fe_{81}Ga_{19}$ ribbon samples fabricated in different rotating speeds of 1 (a), 2 (b), 3 krpm (c) and those of the Ce doped ribbon samples (d), (e), (f), respectively. In the figures, blue arrows indicate Ce rich precipitations. Composition analyzing areas by EDX for $Fe_{81}Ga_{19}$-1 krpm, +$Pr_{0.2}$-1 krpm and +Pr-1 krpm at the free face, (g), (h), (i). Area of +Pr-1k at the roll face (j).

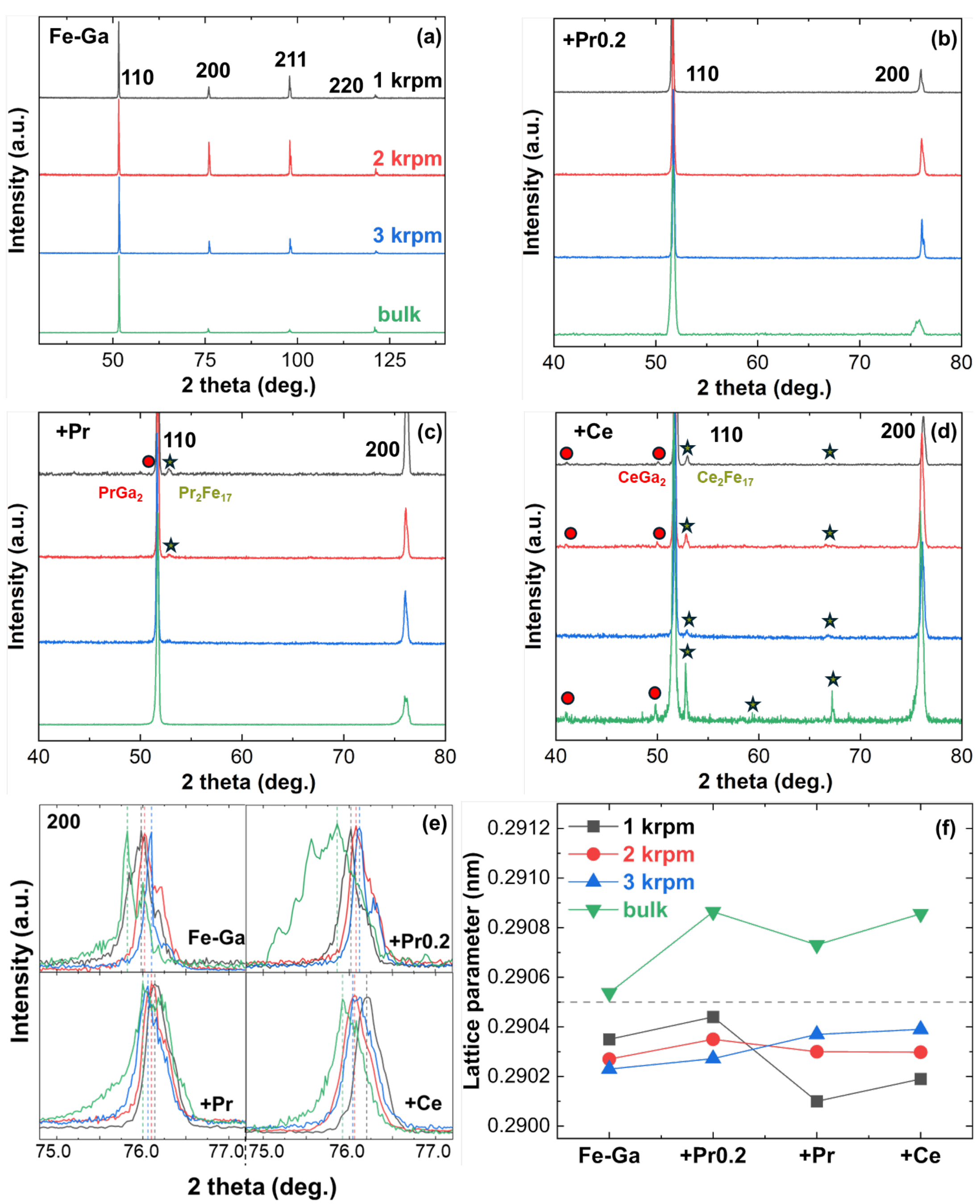

Figure 2 XRD patterns of $Fe_{81}Ga_{19}$ (a), 0.2% Pr (b), 1% Pr (c) and 1% Ce (d) doping sample fabricated in different rotating speeds of 1, 2 and 3 krpm, together with that of the bulk alloy [38]. Expanded scale of 200 diffraction peaks for each system (e). Lattice parameters refined from XRD patterns in all systems (f).

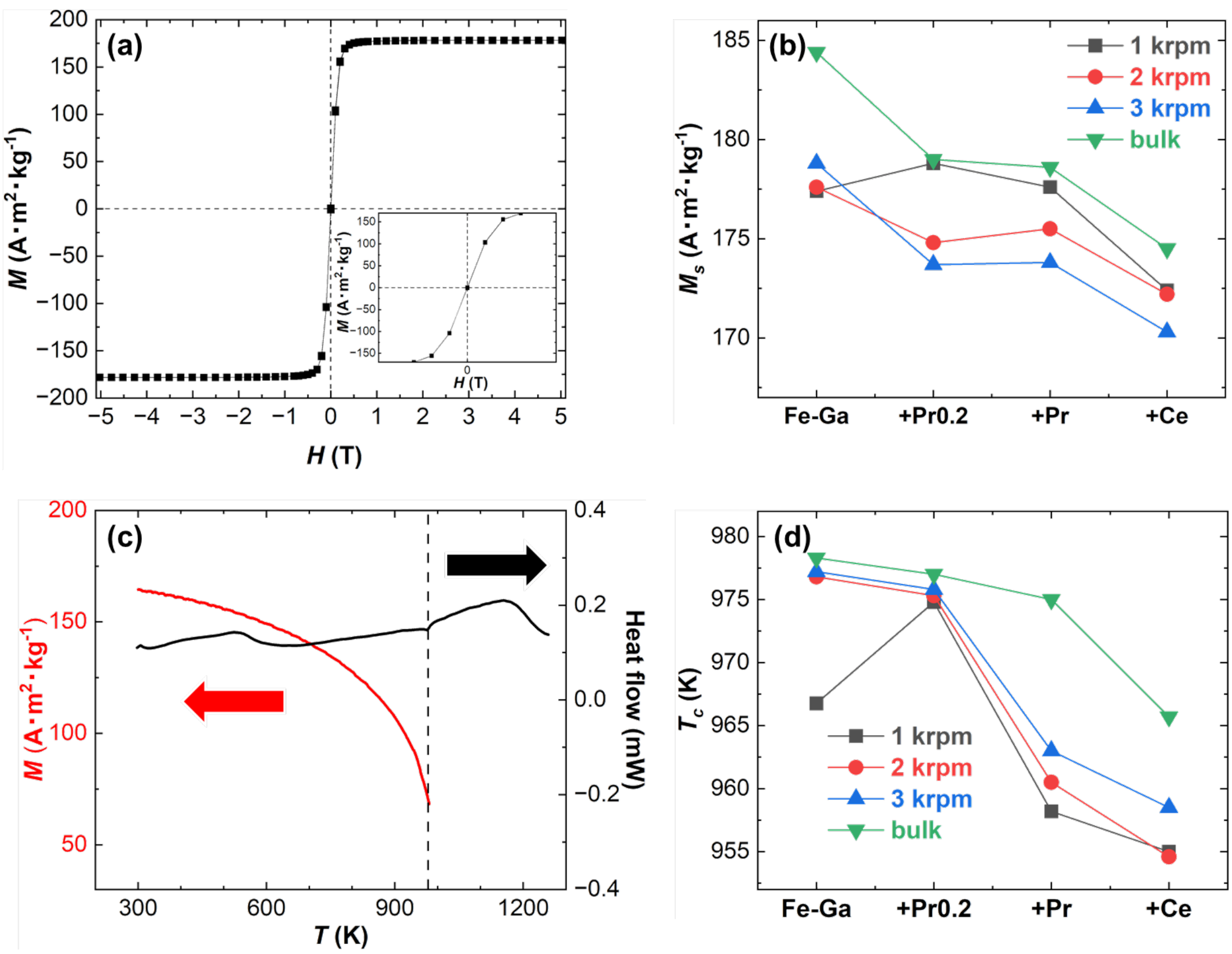

Figure 3 Hysteresis loop of $Fe_{81}Ga_{19}$ 1 krpm ribbon sample measured at 5 K (a)[38] and the value of the saturation magnetization, $M_s$, evaluated by *M*-*H* curves (b). DSC heating curve and temperature dependence of magnetization (*M*-*T*) (c), and the Curie temperature, $T_C$, determined by both DSC and *M*-*T* curves in all the samples (d).

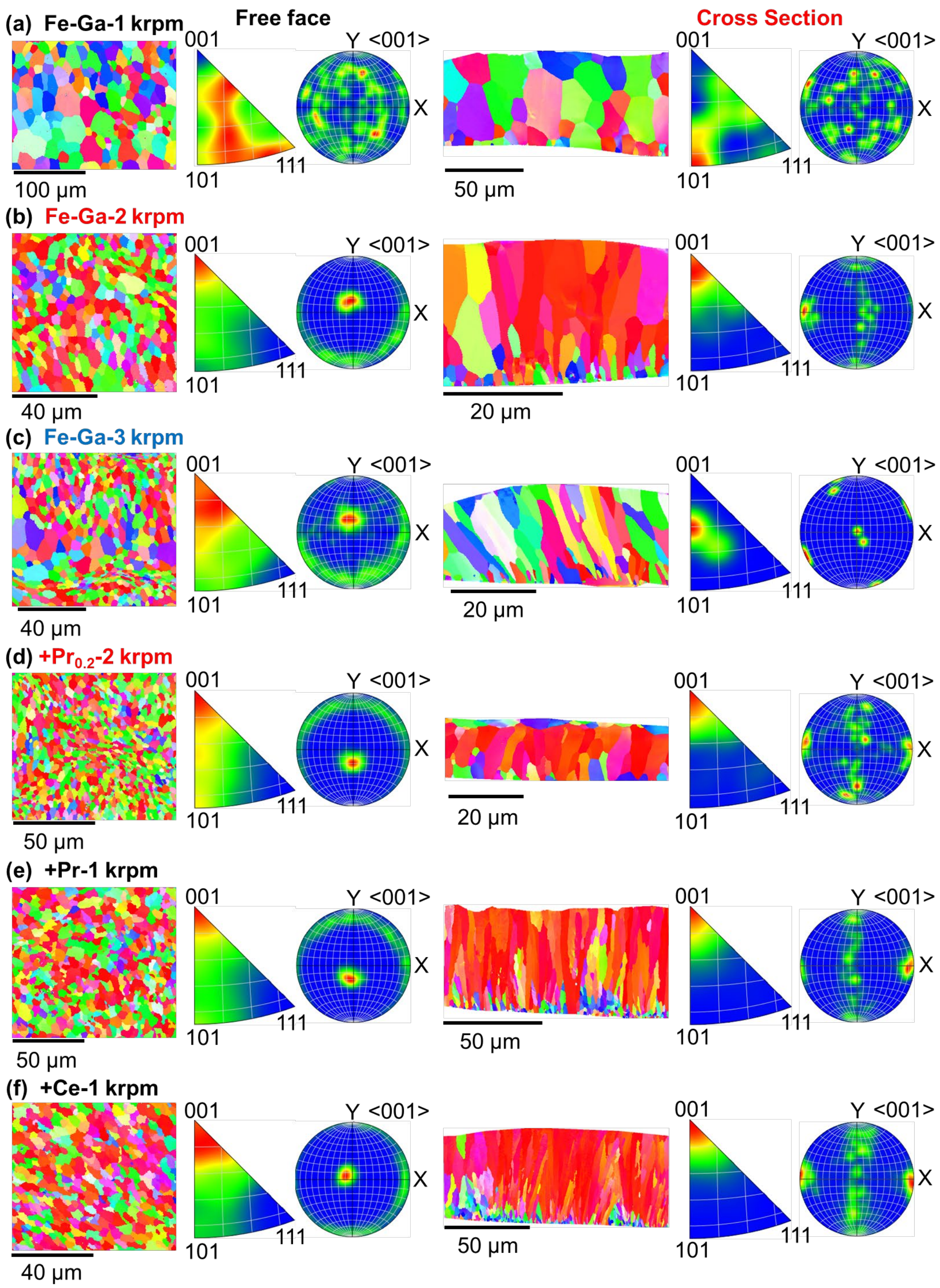

Figure 4 EBSD maps, polar figures, and inverse polar figures observed in both free face and cross section in A1 direction (X axis) for ribbons $Fe_{81}Ga_{19}$ fabricated by in 1, 2 and 3 krpm rotating speeds (a), (b), (c) and the best arranging ribbons of $+Pr_{0.2}$ (d), +Pr (e) and +Ce (f).

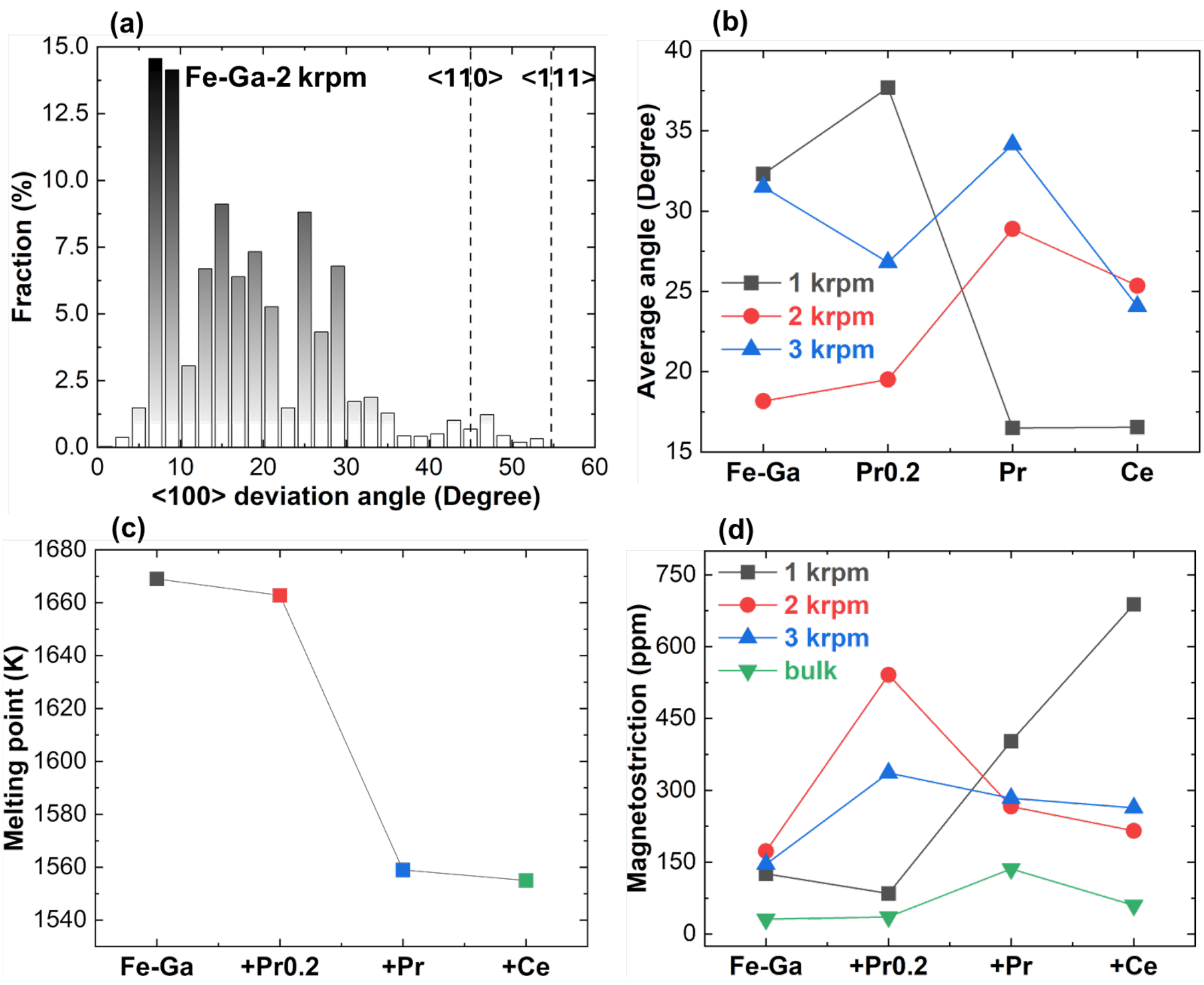

Figure 5 Distribution of azimuth angle difference from <100> for $Fe_{81}Ga_{19}$-2 krpm ribbon (a) and the averaged angle (b). Melting temperature of the mother alloys in each sample (c). Magnetostriction of all specimens, together with that of bulk (d) [78].

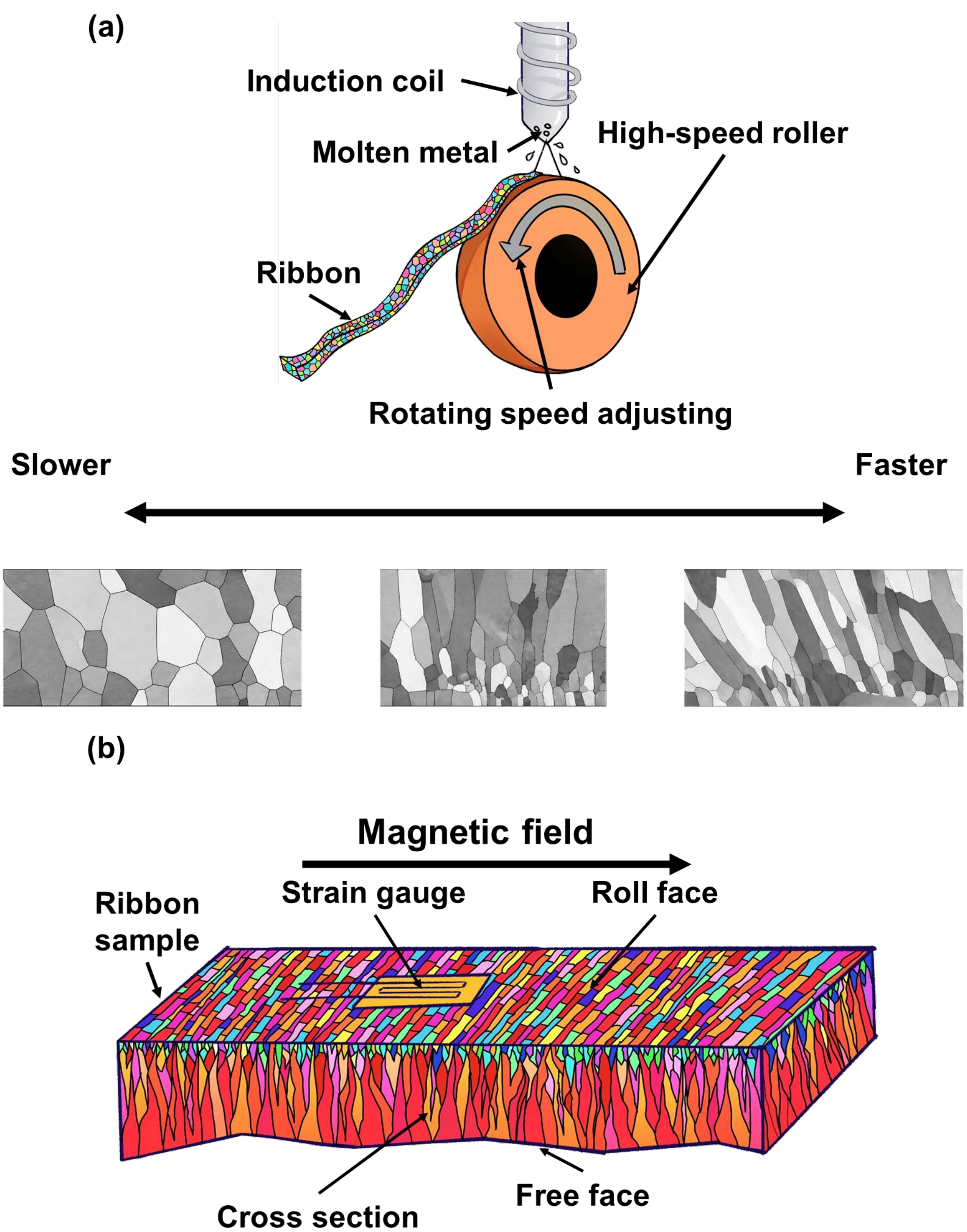

Figure 6 (a) Schematic images of melt-spun method. The texture is controllable with rotating speed adjusting. (b) Schematic view of measurement of magnetostriction of melt-spun ribbon samples. Strain gage is sticked in the roll face and the magnetostriction are evaluated in both parallel and perpendicular direction to the applied magnetic field.

**(g) Free face of $Fe_{81}Ga_{19}$ ribbon (1 krpm)**

| Micro-zones | Elements (at.%) | |
|---|---|---|
| | Fe | Ga |
| Full map | 80.0 | 20.0 |
| Area I | 80.4 | 19.6 |
| Area II | 79.8 | 20.2 |

**(h) Free face of $(Fe_{81}Ga_{19})_{99.8}Pr_{0.2}$ ribbon (1 krpm)**

| Micro-zones | Elements (at.%) | | |
|---|---|---|---|
| | Fe | Ga | Pr |
| Full map | 80.7 | 19.2 | 0.1 |
| Area I | 80.0 | 19.5 | 0.5 |
| Area II | 80.9 | 18.6 | 0.5 |
| Point I | 74.7 | 16.4 | 8.9 |

**(i) Free face of $(Fe_{81}Ga_{19})_{99}Pr_1$ ribbon (1 krpm)**

| Micro-zones | Elements (at.%) | | |
|---|---|---|---|
| | Fe | Ga | Pr |
| Full map | 80.6 | 19.1 | 0.3 |
| Area I | 81.6 | 18.0 | 0.4 |
| Area II | 79.4 | 20.2 | 0.4 |
| Point I | 23.5 | 11.5 | 65.0 |

**(j) Roll face of $(Fe_{81}Ga_{19})_{99}Pr_1$ ribbon (1 krpm)**

| Micro-zones | Elements (at.%) | | |
|---|---|---|---|
| | Fe | Ga | Pr |
| Full map | 80.4 | 19.5 | 0.1 |
| Area I | 81.5 | 18.3 | 0.2 |
| Area II | 81.0 | 18.7 | 0.3 |
| Point I | 73.4 | 13.3 | 13.3 |

Table 1 List of the compositions evaluated by EDX analyses. Alloy compositions correspond with figures 1, (g), (h), (i), (j). The $Fe_{81}Ga_{19}$ ribbon sample showed homogeneous composition but RE rich phases existed in both 0.2% and 1% RE doped samples.

**Supporting information**

# Close correlation between giant magnetostriction and the microstructure in Fe-Ga melt-spun ribbons

Likun Chen[1,2], Mitsutaka Sato[3], Jiahao Han[4], Rie Y. Umetsu[1,4]

The microstructures of $Fe_{81}Ga_{19}$ ribbons observed at roll face and Pr-doped ribbons at free face are shown in figures S1 (a)-(c) and (d)-(f), respectively. Similar to the behavior observed on the free face, the grain size on the roll face of Fe–Ga also becomes finer. A similar trend is observed for the Pr-doped samples, where the grain size on the free face decreases with increasing cooling rate. In addition, some white spherical RE-rich secondary phases are observed of the free face in the 1 at.% Pr-doped samples. As in the Ce-doped samples, the amount of these white spherical secondary phases decreases with increasing rotation speed.

The XRD patterns of the RE-doped samples are presented in figures S2 (a)-(c). The matrix phase of all the samples is identified as the A2 phase. In the 0.2 at.% Pr-doped samples, no extra diffraction peaks are observed because of the low RE concentration. In contrast, additional peaks appear in the 1 at.% Pr- and Ce-doped samples, indicating the formation of RE-containing second phases. The intensity of these extra peaks decreases with increasing cooling rate.
The magnetic hysteresis (*M–H*) curves of all samples are shown in figures S3 (a)-(d). All samples exhibit easily magnetic saturation and relatively high values of the saturation magnetization. The magnetization changes slightly with cooling rate, which would be attributed to variations in the matrix composition. The DSC heating curves are shown in figures S4. A small peak in the DSC curve is assigned to the Curie temperature.

The cross-sectional EBSD maps, polar figures and inverse polar figures of the RE-doped samples at different cooling rates are presented in figures S5 (a)–(c). The results of the specimens with the best controlled texture in each system were indicated and discussed in the main text. Here, the rest of the results are shown. In the $+Pr_{0.2}$-1 krpm sample, most grains exhibit an equiaxed or polygonal morphology, which is attributed to the relatively low cooling rate. In contrast, in the 3 krpm Pr0.2, Pr-, and Ce-doped samples, as well as in the 2 krpm Ce-doped sample, columnar grains with a preferred <100> orientation are clearly observed growing from the bottom of the ribbon. However, grains with <110> and <111> orientations near the free face suppress the continuous growth of the <100>-oriented columnar grains originating from the roll face, thereby reducing the overall fraction of the <100> texture.
Figure S5 (d) indicates DSC heating curves for evaluating melting temperatures, $T_m$, of the four kinds of specimens. It is shown that $T_m$ for the matrix decreased in the RE doped specimens.

The $\lambda$–$H$ curves are shown in figure S6, demonstrating that the ribbons exhibit higher magnetostriction than that of the bulk samples, while magnetic saturation occurs at a relatively higher field. Although some ribbon samples are not fully saturated within the applied magnetic field range, the relative order of magnetostriction among the different samples remains unchanged. To verify the reliability of the magnetostriction measurement, the time dependence of both the applied magnetic field and the magnetostriction response is also shown in figure S6(e). The saturation magnetostriction remains unchanged over three field-cycling measurements, confirming the reproducibility of the experiments.

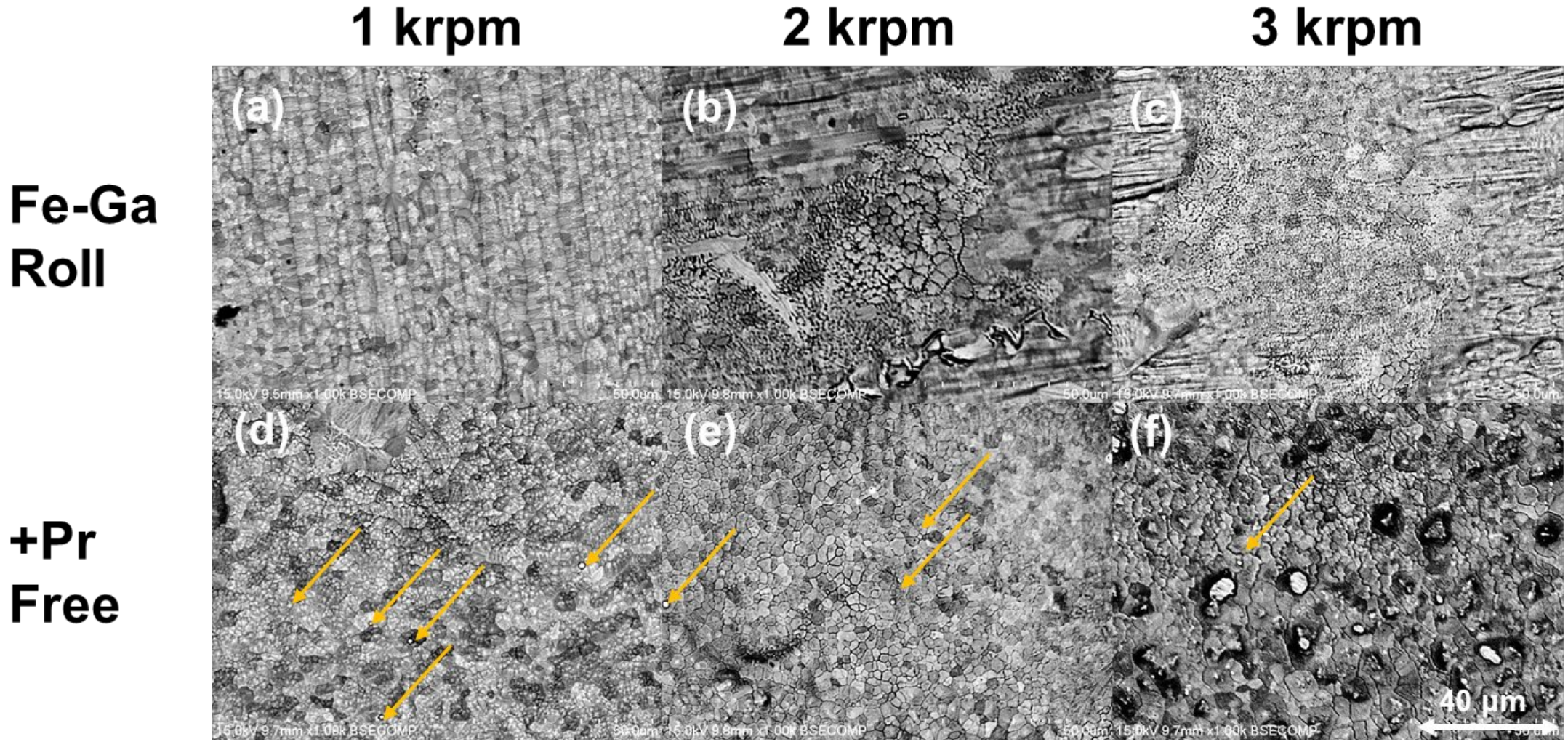

Figure S1. Microstructures of the roll face of $Fe_{81}Ga_{19}$ ribbons fabricated by 1, 2 and 3 krpm rotating speeds, (a), (b), (c), respectively, and those at the free face of +Pr sample (1 at.% of Pr was doped) (c), (d), (e). Here, the amount of RE-rich secondary phase decreases with cooling rates increasing as indicated by the yellow arrows.

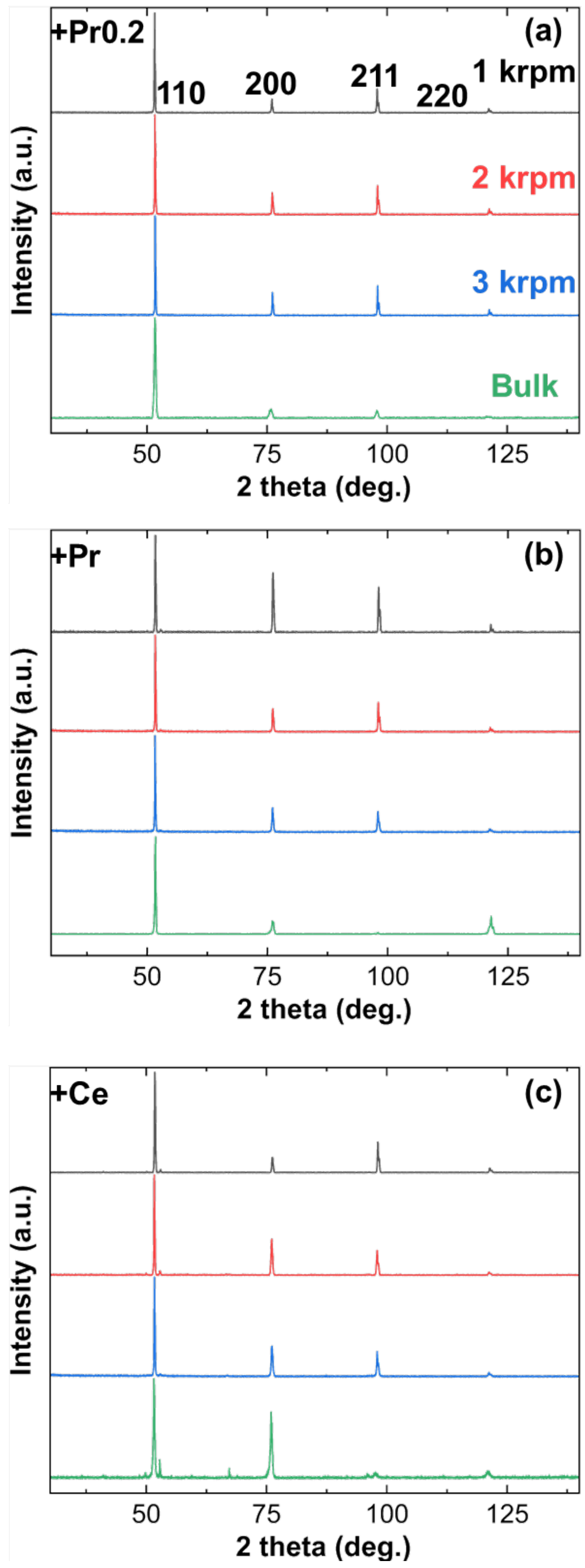

Figure S2. XRD patterns of $+Pr_{0.2}$ (a), +Pr (b) and +Ce (c) addition to that of bulk reported before[78].

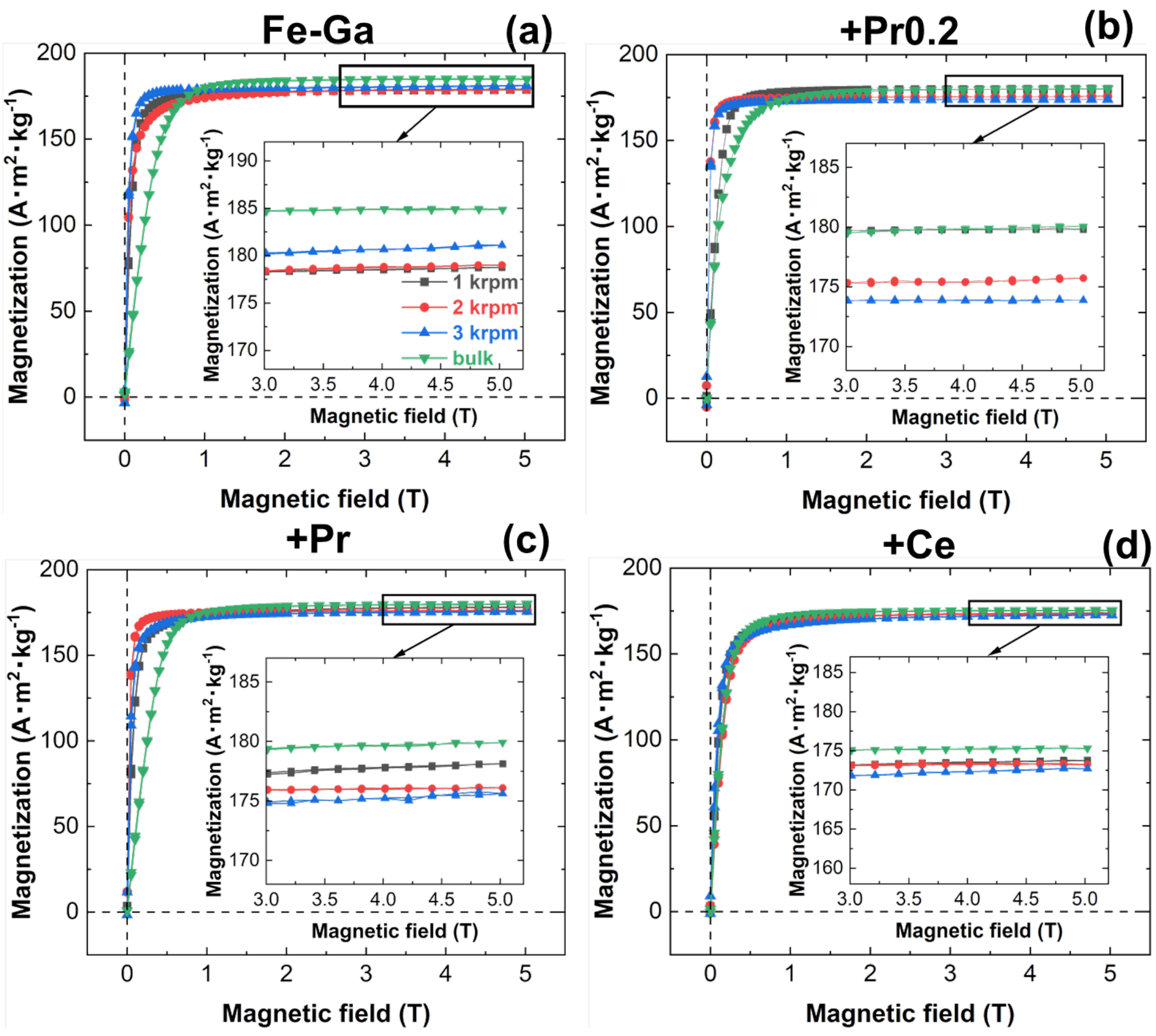

Figure S3. *M*-*H* curves measured at 5 K of all ribbon and bulk samples of $Fe_{81}Ga_{19}$ (a), $+Pr_{0.2}$ (b), +Pr (c) and +Ce (d). The bulk data reported before[78].

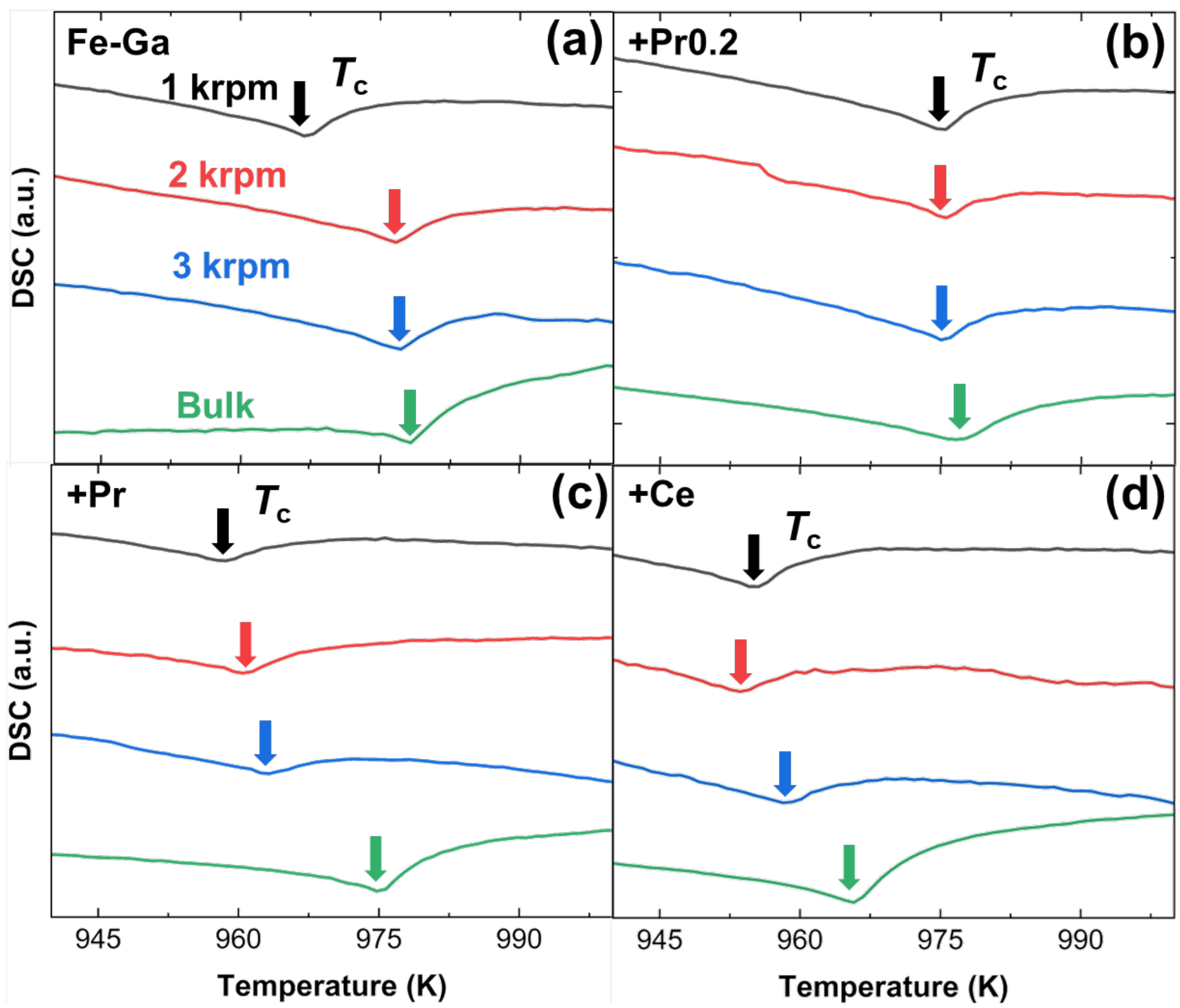

Figure S4. DSC heating curves for Curie temperature evaluation of $Fe_{81}Ga_{19}$ (a), $+Pr_{0.2}$ (b), +Pr (c) and +Ce (d), together with the data of bulk [78].

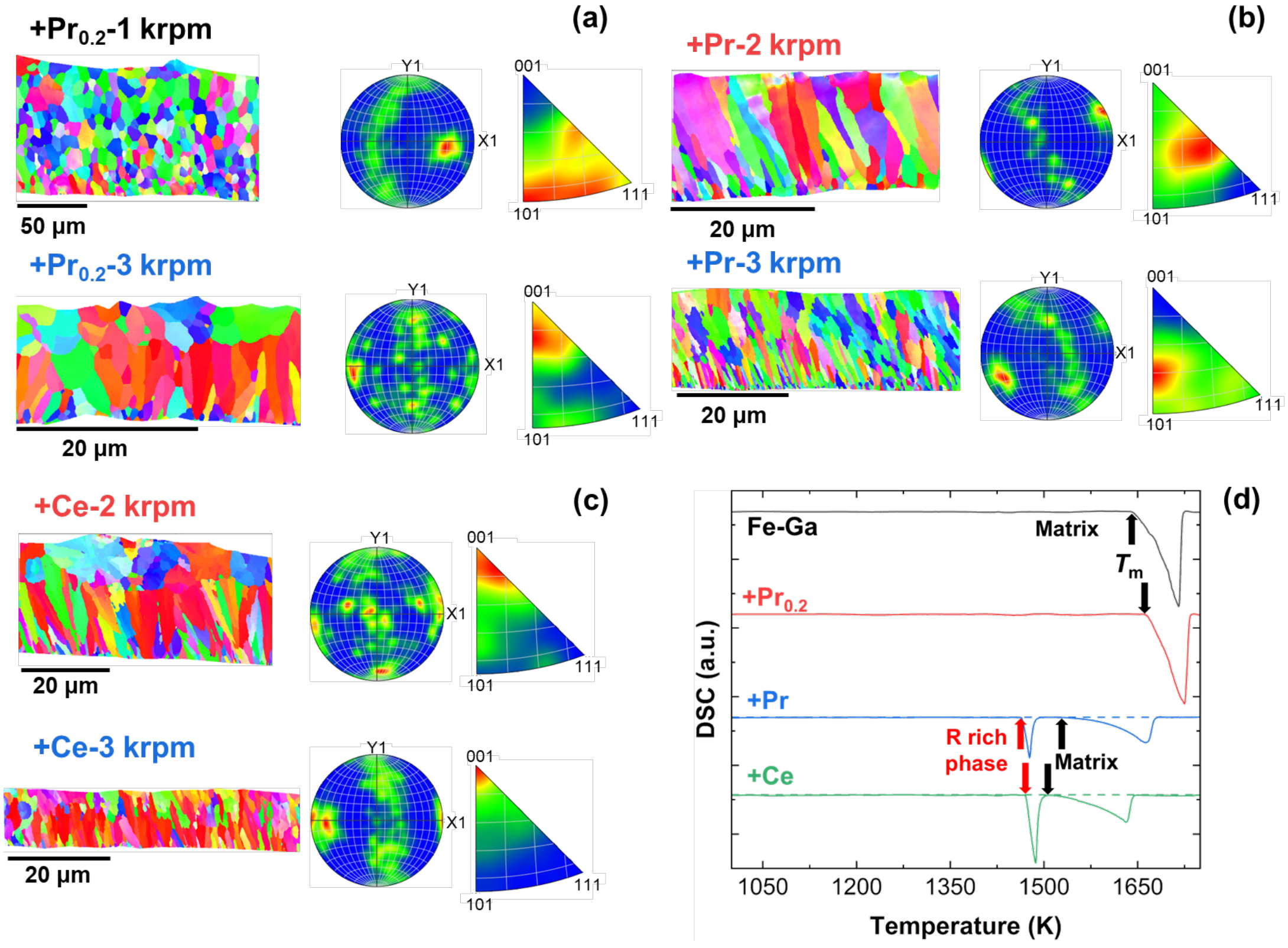

Figure S5. EBSD images, polar figures and inverse polar figures of $+Pr_{0.2}$(a), +Pr (b) and +Ce (c). DSC heating curves for evaluating the melting temperatures of $Fe_{81}Ga_{19}$, and the doped series.

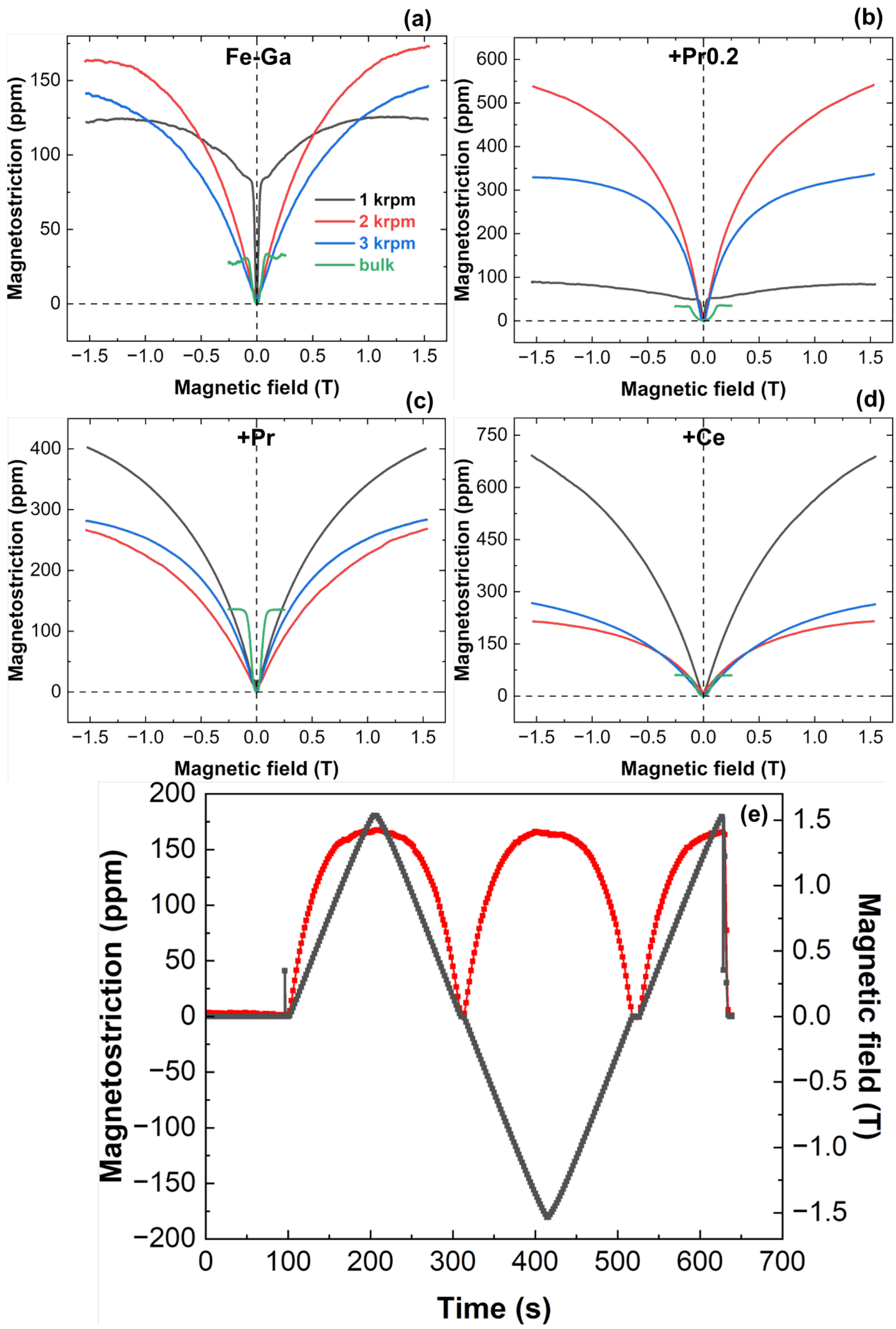

Figure S6. All magnetostriction-magnetic field curves in $Fe_{81}Ga_{19}$ (a), $+Pr_{0.2}$, (b), +Pr (c) and +Ce (d), together with the data of bulk[78]. Dynamic response of magnetostriction of $Fe_{81}Ga_{19}$-2 krpm specimen under applied magnetic field cycling (e).